\RequirePackage{etex}
\documentclass[AMA,STIX2COL]{MRM}
\articletype{Full Paper}%

\begin{document}

\title{Human brain high-resolution diffusion MRI with optimized slice-by-slice $B_0$ field shimming in head-only high-performance gradient MRI systems \protect}

\author[1]{Patricia Lan}{\orcid{0000-0003-1763-9540}}

\author[2]{Sherry S. Huang}{\orcid{0000-0003-0614-6628}}

\author[3]{Chitresh Bhushan}{\orcid{0000-0002-2935-2515}}

\author[4]{Xinzeng Wang}{\orcid{0000-0002-0415-7246}}

\author[3]{Seung-Kyun Lee}{\orcid{0000-0001-7625-3141}}

\author[5,6]{Raymond Y. Huang}{\orcid{0000-0001-7661-797X}}

\author[7]{Jerome J. Maller}{\orcid{0000-0003-4685-1508}}

\author[8]{Jennifer A. McNab}{\orcid{0000-0002-1763-0792}}

\author[3]{Ante Zhu}{\orcid{0000-0003-2648-0260}}

\authormark{Lan \textsc{et al}}

\address[1]{\orgdiv{MR Clinical Solutions \& Research Collaborations}, \orgname{GE HealthCare}, \orgaddress{\state{Menlo Park, CA}, \country{USA}}}

\address[2]{\orgdiv{Science and Technology Office}, \orgname{GE HealthCare}, \orgaddress{\state{Royal Oak, Michigan}, \country{USA}}}

\address[3]{\orgdiv{Technology \& Innovation Center}, \orgname{GE HealthCare}, \orgaddress{\state{Niskayuna, New York}, \country{USA}}}

\address[4]{\orgdiv{MR Clinical Solutions \& Research Collaborations}, \orgname{GE HealthCare}, \orgaddress{\state{Houston, Texas}, \country{USA}}}

\address[5]{\orgdiv{Department of Radiology}, \orgname{Brigham and Women's Hospital}, \orgaddress{\state{Boston, Massachusetts}, \country{USA}}}

\address[6]{\orgname{Harvard Medical School}, \orgaddress{\state{Boston, Massachusetts}, \country{USA}}}

\address[7]{\orgdiv{Department of Radiology}, \orgname{Iowa University}, \orgaddress{\state{Iowa City, Iowa}, \country{USA}}}

\address[8]{\orgdiv{Department of Radiology}, \orgname{Stanford University}, \orgaddress{\state{Stanford, California}, \country{USA}}}

\corres{Ante Zhu, GE HealthCare Technology \& Innovation Center, One Research Circle, MR 135. Niskayuna 12309, NY, USA. \email{ante.zhu@gehealthcare.com}}

\finfo{NIH U01EB034313}

%% Abstract word limit: 250

\abstract{
\section{Purpose} To propose a brain tissue-selective, optimized slice-by-slice $B_0$ field shimming for high-resolution brain diffusion MRI. 
\section{Methods} We proposed to incorporate actual gradient fields of X, Y, and Z gradient coils in the calculation of the shimming coefficients in dynamic slice-by-slice $B_0$ field shimming to minimize $B_0$ field inhomogeneity (i.e., $\Delta B_0$) in deep-learning segmented brain tissues. Diffusion MRI with oscillating gradient spin echo (OGSE) at 55 Hz and pulsed gradient spin echo (PGSE) (approximated at 0 Hz) were obtained in phantoms and healthy volunteers using a head-only high-performance gradient 3T MRI system. In each diffusion MRI acquisition, standard static volumetric shimming and the proposed shimming method were applied separately, and mean/axial/radial diffusivities (MD/AD/RD) and fractional anisotropy (FA) were estimated.
\section{Results} In phantom, the root-mean-square of $\Delta B_0$ in areas with high gradient nonlinearity was reduced by 7 Hz when incorporating actual gradient field in dynamic shimming. Compared to static shimming, dynamic shimming reduced root-mean-square of voxel displacement of each slice by a maximum of 5-10 voxels in single-shot EPI acquisition at 1-2 mm in-plane resolution in phantom, and a maximum of 3 voxels in human brains. Improved accuracy of MD/AD/RD/FA in the superior region of the brain, brainstem, and cerebellum were observed by applying dynamic shimming and/or two-shot EPI acquisition. MD(55 Hz)-MD(0 Hz) showed higher values in $T_2$ FSE hypo-intensity region by applying dynamic shimming.
\section{Conclusion} Diffusion MRI with brain tissue-selective, dynamic slice-by-slice $B_0$ effectively improves the accuracy of diffusivity characterization in high-resolution images.}


%\jnlcitation{\cname{%
%\author{Lan P.},
%\author{Huang SS}, 
%\author{Bhushan C},
%\author{Wang X},
%\author{Lee S.},
%\author{Huang RY},
%\author{Maller J},
%\author{McNab JA},
%and
%\author{Zhu A.}} (\cyear{2025}), 
%\ctitle{Human brain high-resolution diffusion MRI with optimized slice-by-slice $B_0$ field shimming in head-only high-performance gradient MRI systems}, \cjournal{Magn. Reson. Med.}, \cvol{2025;00:1--12}.}

\maketitle

%\footnotetext{\textbf{Abbreviations:}~\hbox{OGSE,~Oscillating~gradient~spin~echo}{\hfill\break}presenting~cells;}

\clearpage 

\section{Introduction}\label{sec1}

High-resolution advanced diffusion MRI methods that provide novel image contrast sensitive to tissue mesoscopic and microscopic structures \cite{Zhu2023MRM, Corey2015, Michael2022, Michael2024, CHAKWIZIRA2023NI, Ramos-Llorden2025, Zhou2025} have emerged to reveal pathological changes in human in vivo studies. These techniques are largely advanced by high-performance gradient human MRI systems. For example, small-bore head-only gradient MRI systems can simultaneously achieve maximum gradient amplitudes of 200-500 mT/m, slew rates of 750-900 T/m/s, \cite{Setsompop2013, Weiger2018MRM, foo_highly_2020, Feinberg2023, Ramos-Llorden2025} and 3-5 times higher peripheral nerve stimulation threshold. \cite{Tan2020, Feinberg2025} These specifications shorten diffusion encoding duration and the echo time of pulsed gradient spin echo (PGSE), resulting in increased signal-to-noise ratio and allowing efficient imaging of high-resolution diffusion MRI. \cite{Feinberg2023, Ramos-Llorden2025, Zhu2024} Furthermore, high-gradient performance enables advanced diffusion encoding including oscillating gradient spin echo (OGSE) that achieves a short diffusion time and high sensitivity to short tissue length scale of $\sim 10 \mu m$. \cite{Zhu2023MRM, Dai2023NI, Michael2024, Hao2025} For example, high-resolution OGSE diffusion MRI has demonstrated clear contrast between the molecular layer and granule cell layers in the adult mouse cerebellum.\cite{Wu2014, Wu2018} This technique is now becoming translatable to human research and clinical studies. 

High-quality diffusion MR imaging is a necessity for accurate characterization of tissue microstructures and anatomical connectivities. However, diffusion MRI with echo planar imaging (EPI) or spiral acquisitions suffers from image distortion or blurring due to $B_0$ field inhomogeneity ($\Delta B_0$), affecting the diagnostic performance, tractography, and quantitative diffusion measurements.\cite{irfanoglu_effects_2012, brun_diffusion_2019} Thus, achieving high-resolution diffusion MRI is challenging due to increased image distortion and blurring at high spatial resolution. Different approaches have been applied to reduce image distortion or blurring in EPI and spiral acquisitions, including (1) shortening the readout via faster slew rate or undersampling which rely on gradient performance or multi-channel radiofrequency receive coil; (2) using multi-shot (or segmented) readout but at the cost of at least 1.5X scan time; (3) improving $B_0$ field homogeneity or measuring and correcting for $B_0$ field inhomogeneity. Here we focus on the reduction of $\Delta B_0$, which may not require special hardware or extra scan time. 

$B_0$ field shimming tunes center frequency, linear gradient fields from the X, Y, Z gradient coils, and/or high-order gradient fields from high-order shim coils to counteract $\Delta B_0$ in a defined region-of-interest (ROI) to zero. Conventional static $B_0$ shimming minimizes $\Delta B_0$ in a 3D volume. Dynamic slice-by-slice $B_0$ field shimming has been developed and shown to further reduce the root-mean-square of $\Delta B_0$ in each slice, demonstrating reduced geometric distortion in 2D EPI acquisitions in the head and neck,\cite{Walter2017, Haisam2019} and body imaging.\cite{lee2014, Qiu2021, McElroy2021, Tollefson2023, Liu2025}

Dynamic $B_0$ field shimming with zeroth and first order leverage the center frequency and X, Y, Z gradient coils to generate zeroth and first order magnetic field for shimming. This approach is accessible to any MRI system, unlike static high-order $B_0$ shimming, which requires additional shim coils and drivers. Zeroth order $B_0$ shimming is tuned by adjusting the center frequency of the radiofrequency excitation and receive demodulation. First-order $B_0$ shimming involves generating a magnetic field by applying currents to the X, Y, and Z gradient coils. The resulting shimming field is calculated as the product of the $B_0$ shimming coefficients (i.e., the currents applied to each coil) and the magnetic field produced by each gradient coil per unit current. The magnetic field of each gradient coil is determined by its electromagnetic design and has been typically approximated as a linear gradient field. However, the small-bore head-only X, Y, Z gradient coils present relatively high gradient nonlinearity \cite{Setsompop2013, Tan2013MRM, Weiger2018MRM, foo_highly_2020, Feinberg2023, Zhu2025} in the 20-26 cm diameter sphere volume, compared to the whole-body gradient coils. In whole-body MRI systems, the effect of gradient nonlinearity on the ROI can be mitigated by moving the ROI to the iso-center where gradient nonlinearity is negligible. In head-only systems, the small bore size of 33-44 cm diameter prohibits the inferior brain regions, including the cerebellum, to be scanned at the iso-center. Therefore, ignoring gradient nonlinearity of the X, Y, Z gradient coils in the calculation of $B_0$ shimming coefficients may result in suboptimal $\Delta B_0$ shimming, especially in the inferior brain regions in head-only gradient MRI systems.

Furthermore, targeting static $B_0$ field shimming for only brain tissues has been shown to more effectively reduce $\Delta B_0$ and image distortion in EPI acquisitions in brain tissues,\cite{Xu2023} compared to conventional static $B_0$ field shimming that targets the entire head. However, automated brain tissue-selective dynamic slice-by-slice $B_0$ field shimming has not been investigated.

%and spinal cord,\cite{kaptan_automated_2022}

The purpose of this study was first to assess the effect of gradient nonlinearity of X, Y, Z gradient coils on dynamic slice-by-slice $B_0$ field shimming in head-only high-performance gradient human 3T MRI systems. Furthermore, we propose to optimize the shimming in the brain by leveraging fast, automated, deep-learning-based brain extraction to define the shimming ROI and incorporating the actual gradient field of the X, Y, Z gradient coils in the calculation of $B_0$ field shimming coefficients. The technique was performed in diffusion MRI of phantoms and healthy volunteers with 2D EPI acquisitions in a human head-only high-performance gradient 3T MRI system.\cite{foo_highly_2020} Whole-brain single-shot EPI and two-shot EPI with both static and dynamic shimming were imaged to evaluate the image contrast of OGSE and PGSE diffusion in comparison with anatomical images.

\section{Methods}\label{sec2}
\subsection{$B_0$ Field Shimming Algorithms}

Here we consider dynamic 2D slice-by-slice $B_0$ shimming with zeroth and first order. Although the dynamic 2D high-order shimming is interesting, the settling time of high-order coils can be a few seconds, prohibiting change of currents for each slice in a TR on the order of milliseconds. $B_0$ shimming coefficients are calculated to minimize the root-mean-square of $\Delta B_0$, to be called $\Delta B_{rms}$, in all \textit{(N)} voxels in an ROI:  

\begin{equation} \label{eq1}
[\Delta f, \alpha_{X}, \alpha_{Y}, \alpha_{Z}] = \arg \min \Delta B_{rms}
\end{equation}

\begin{equation} \label{eq2}
\Delta B_{rms}^2 = \frac{1}{N} \sum_{\overrightarrow v \in ROI}  \Bigg|\Delta B_0(\overrightarrow v)-\Delta f - \sum_{j=X,Y,Z}g_j(\overrightarrow v, \alpha_j) \bigg|^2
\end{equation}

\noindent where $\Delta f$ is the applied center frequency for $B_0$ field shimming; $\alpha_{X}$, $\alpha_{Y}$, and $\alpha_{Z}$ are the currents to the three physical gradient coils along the left-right, anterior-posterior, and inferior-superior directions for $B_0$ field shimming; $\sum g$ is the total generated gradient field in each voxel at the location $\overrightarrow{v}$=(x,y,z).

Conventionally, the gradient fields along all the three gradient axes are approximated as linear. Therefore, the second summation in Eq. \ref{eq2} becomes:

\begin{equation} \label{eq3}
\sum_{j=X,Y,Z}  g_i(\overrightarrow v, \alpha_j) = \alpha_X \cdot G_X \cdot x + \alpha_Y \cdot G_Y \cdot y + \alpha_Z \cdot G_Z \cdot z.
\end{equation}

\noindent where $G_{X}$, $G_{Y}$, and $G_{Z}$ are spatially independent, linear gradient field with amplitude generated by applying unit current, i.e., 1 Ampere, to each of the gradient coils.

\begin{figure*}%[t]
\centerline{\includegraphics[width=500pt,height=19pc]{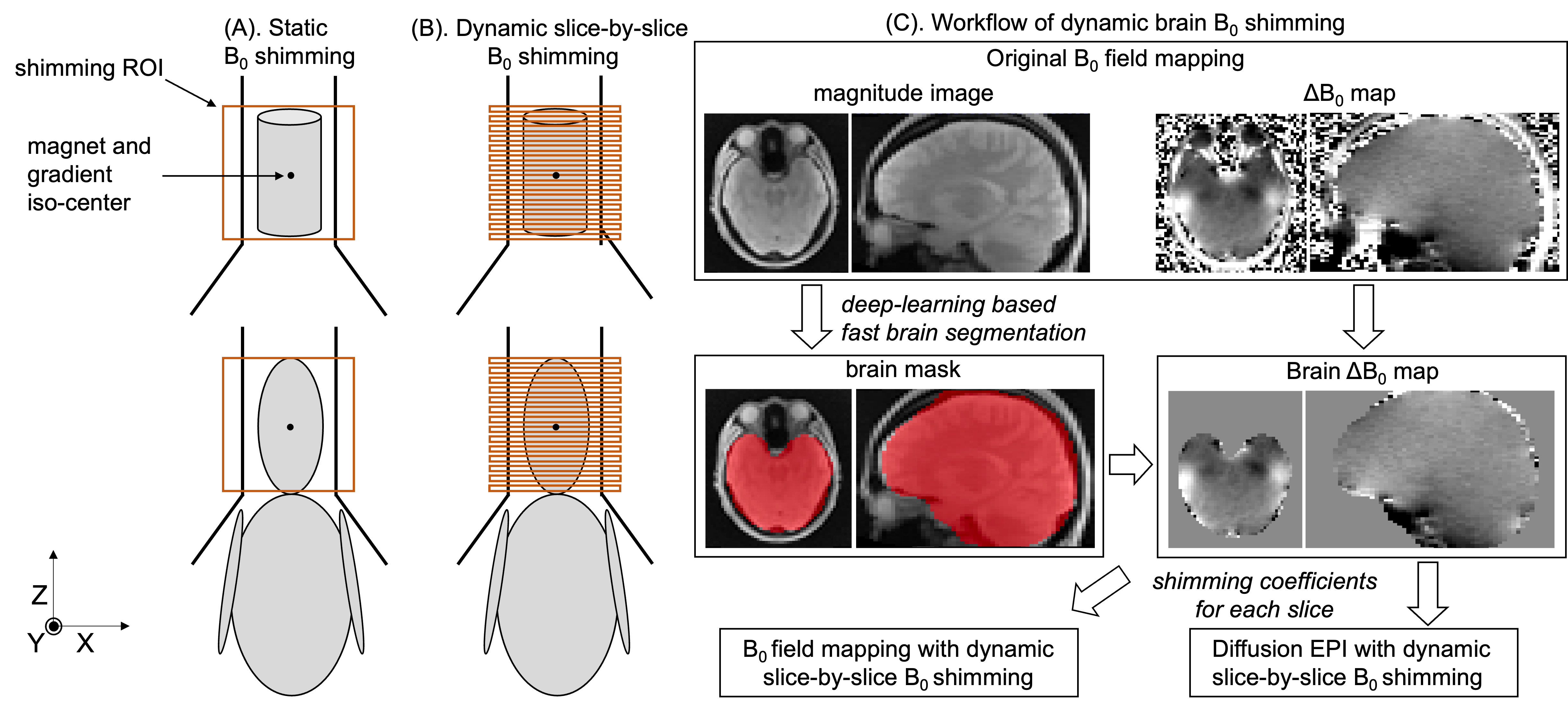}}
\caption{Imaging setup in a head-only high-performance gradient MRI systems for phantoms (top) and human volunteers (bottom) using (A) static and (B) dynamic shimming. (C) Workflow of dynamic $B_0$ field shimming, where an axial $B_0$ field map is obtained, and the brain is segmented from the magnitude images using a deep-learning based algorithm. The resulting masked field map is used to calculate the slice-specific shimming coefficients, which are then applied during a second $B_0$ field mapping sequence and diffusion EPI sequences.\label{fig1}}
\end{figure*}

\begin{table*}%[t]
\centerline{\includegraphics[width=500pt,height=21pc]{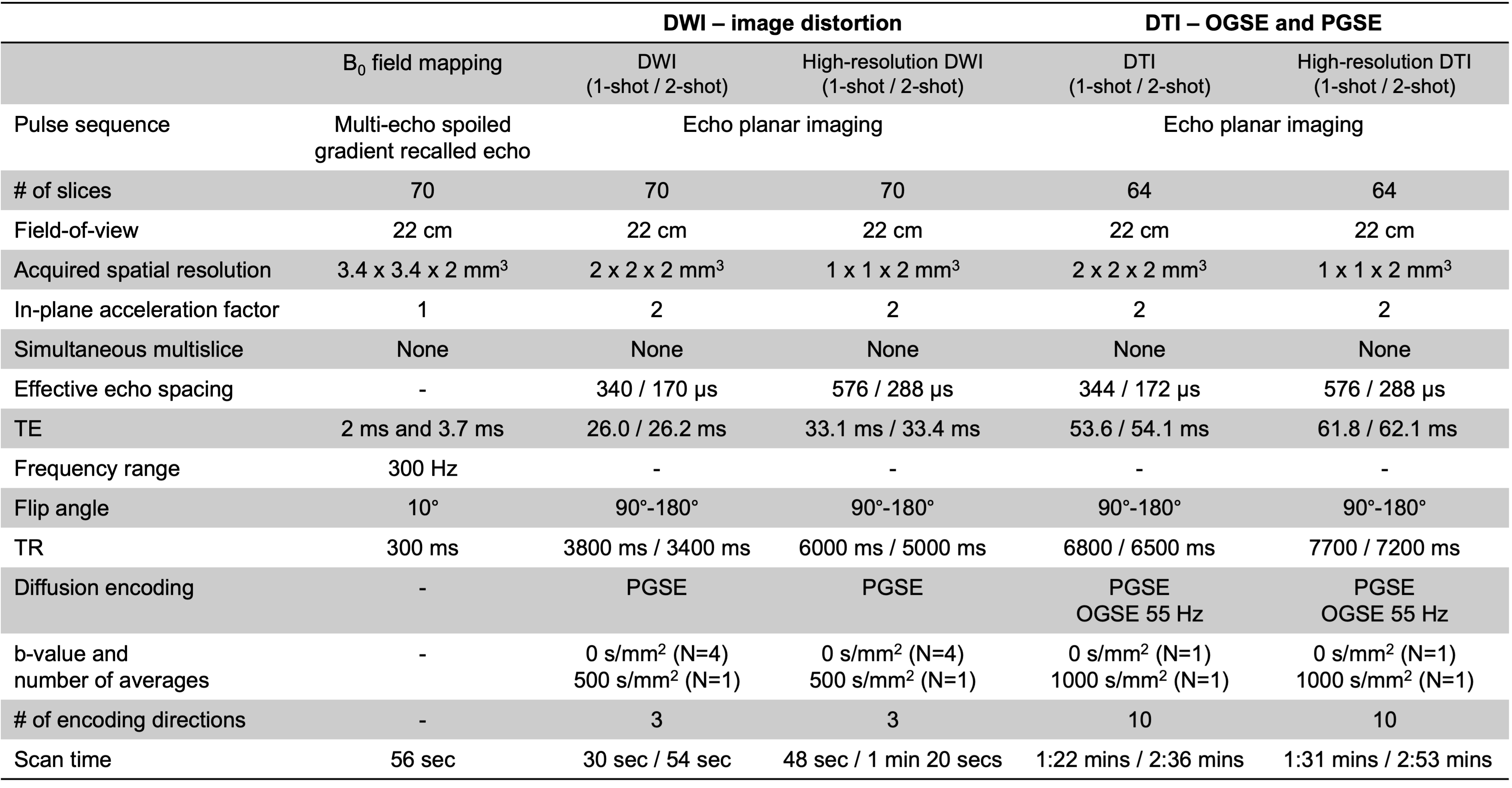}}
\caption{Imaging parameters in an investigational MAGNUS head-only gradient coil operating at maximum performance of 300 mT/m and 750 T/m/s.\label{table1}}
\end{table*}

In the presence of gradient nonlinearity, the actual gradient of each voxel deviates from linear gradient field. In general, Eq. \ref{eq3} becomes 

\begin{equation} \label{eq4}
\begin{aligned}
\sum_{j=X,Y,Z} g_j({\overrightarrow v, \alpha_j}) = \\
\alpha_X \cdot g_{X}(x,y,z) + \alpha_Y \cdot g_{Y}(x,y,z) + \alpha_Z \cdot g_{Z}(x,y,z).
\end{aligned}
\end{equation}

\noindent where \textit{$g_X()$}, $g_Y()$, $g_Z()$ are now functions of space that define the static magnetic field produced by 1 Ampere of current to each of the gradient coils.

To evaluate the effect of gradient nonlinearity on $B_0$ field shimming, two schemes were implemented to minimize the $\Delta B_{rms}$ within a ROI, including: (1) dynamic 2D slice-by-slice shimming assuming linear gradient fields; (2) dynamic 2D slice-by-slice shimming using actual gradient fields. The estimation of $B_0$ shimming coefficients was obtained by using the "fminsearch" function from MATLAB (Mathworks, Natick, MA, USA). 

\begin{figure*}
\centerline{\includegraphics[width=500pt,height=36pc]{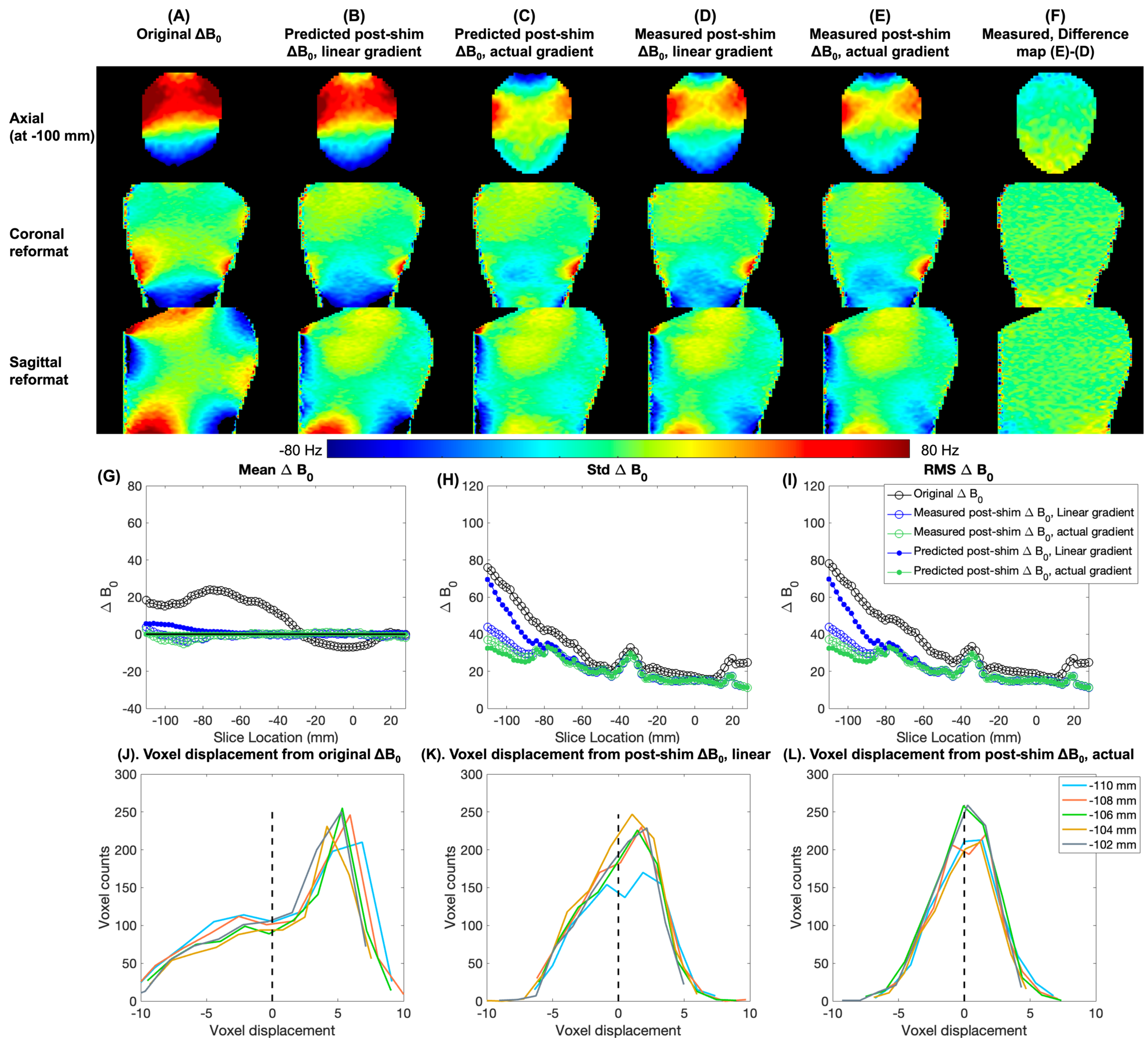}}
\caption{$B_0$ field map analysis of a brain-shaped phantom. (A-F) Axial, coronal reformat, and sagittal reformat of the original $\Delta B_0$, predicted post-shim $\Delta B_0$ using linear/actual gradients, measured post-shim $\Delta B_0$ using linear/actual gradients, and the difference between the measured post-shim $\Delta B_0$ with linear and actual gradient fields. (G-I) Mean, standard deviation, and root-mean-square of $\Delta B_0$ in each slice. Voxel displacements were calculated using an echo spacing of 344 $\mu s$ and single-shot EPI acquisition in the presence of three $B_0$ maps, i.e., (J) the original $\Delta B_0$, (K) measured post-shim $\Delta B_0$ using linear gradients, (L) measured post-shim $\Delta B_0$ using actual gradients. \label{fig2}}
\end{figure*}

\begin{figure*}
\centerline{\includegraphics[width=500pt,height=36pc]{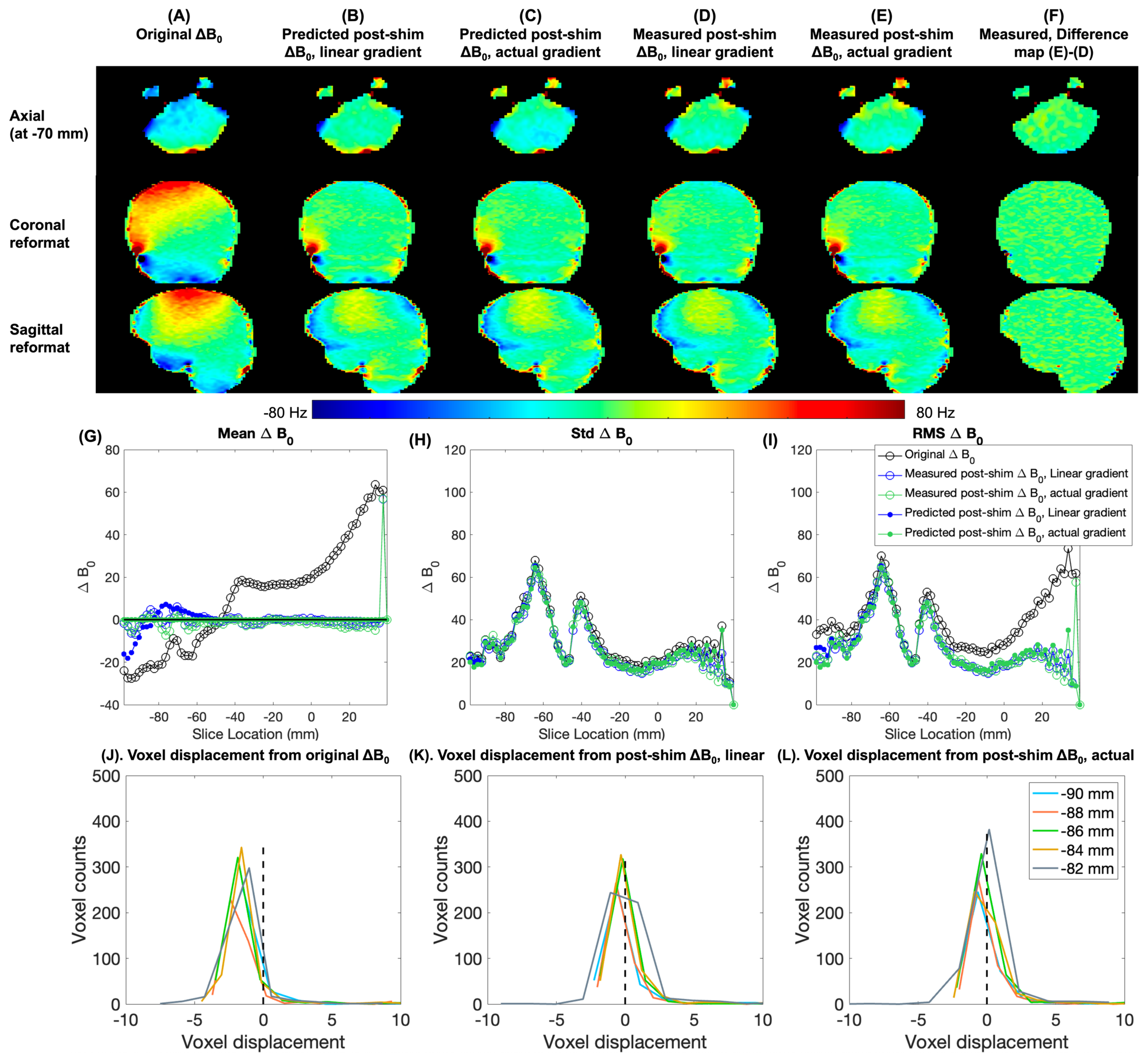}}
\caption{$B_0$ field map analysis of a healthy volunteer's brain (i.e., Subject \#1). (A-F) Axial, coronal reformat, and sagittal reformat of the original $\Delta B_0$, predicted post-shim $\Delta B_0$ using linear/actual gradients, measured post-shim $\Delta B_0$ using linear/actual gradients, and the difference between the measured post-shim $\Delta B_0$ with linear and actual gradient fields. (G-I) Mean, standard deviation, and root-mean-square of $\Delta B_0$ in each slice of brain tissues. Voxel displacements were calculated using an echo spacing of 344 $\mu s$ and single-shot EPI acquisition in the presence of three $B_0$ maps, i.e., (J) the original $\Delta B_0$, (K) measured post-shim $\Delta B_0$ using linear gradients, (L) measured post-shim $\Delta B_0$ using actual gradients. \label{fig3}}
\end{figure*}

\begin{figure*}
\centerline{\includegraphics[width=420pt,height=36.75pc]{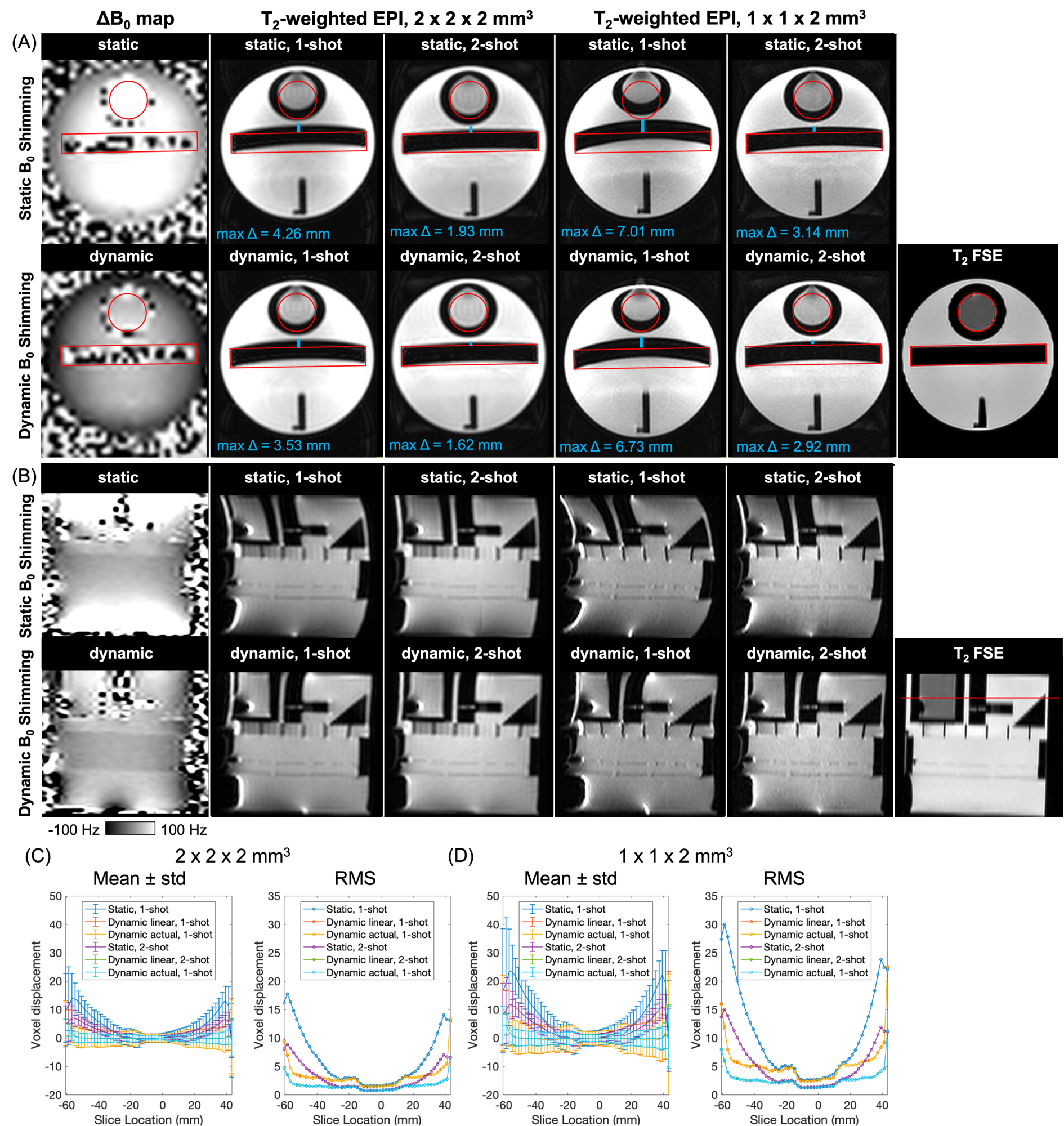}}
\caption{$B_0$ map and $T_2$-weighted EPI at 2 x 2 x 2 $mm^3$ resolution and 1 x 1 x 2 $mm^3$ resolution, and $T_2$ FSE of representative slices of the mini-ACR phantom in axial (A) and sagittal (B) views. The axial slice was noted by the red line in the sagittal $T_2$ FSE image. $T_2$-weighted EPI with dynamic shimming was acquired using the actual gradient field in the calculation of shimming coefficients. The blue lines indicate the maximum voxel displacement of the hollow rectangle compared to $T_2$ FSE, outlined in red. The voxel displacement in the 2 x 2 x 2 $mm^3$ EPI (C) and 1 x 1 x 2 $mm^3$ EPI (D) was calculated based on $\Delta B_0$ and EPI acquisition parameters in Table \ref{table1}. \label{fig4}}
\end{figure*}

\begin{figure*}
\centerline{\includegraphics[width=420pt,height=36.75pc]{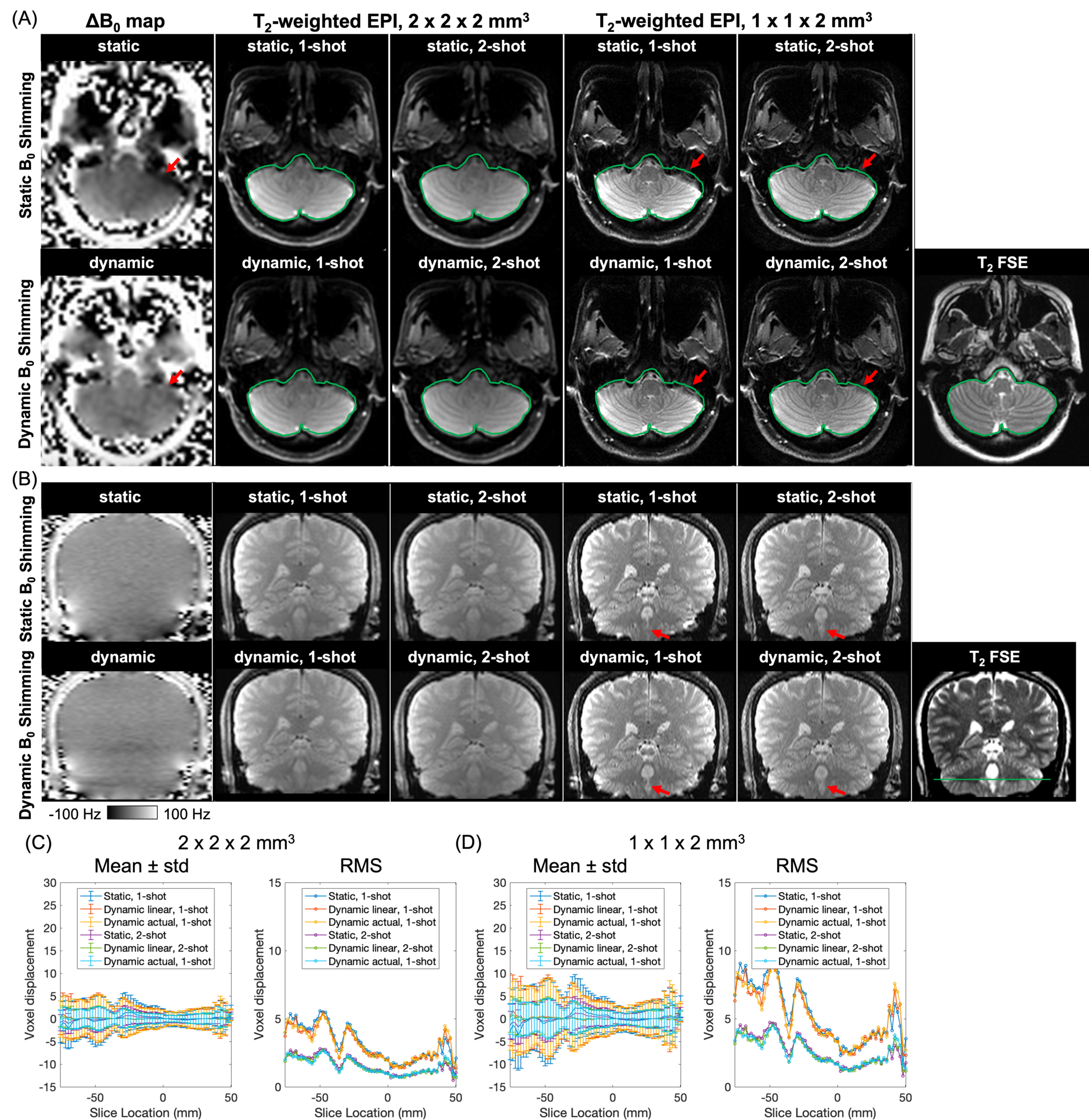}}
\caption{$B_0$ map, $T_2$-weighted EPI at 2 x 2 x 2 $mm^3$ resolution and 1 x 1 x 2 $mm^3$ resolution, and $T_2$ FSE of representative slices of a healthy volunteer's brain (i.e., Subject \#2) in axial (A) and sagittal (B) views. The axial slice was noted by the green line in the sagittal $T_2$ FSE image. $T_2$-weighted EPI with dynamic shimming was acquired using the actual gradient field in the calculation of shimming coefficients. The voxel displacement in the 2 x 2 x 2 $mm^3$ EPI (C) and 1 x 1 x 2 $mm^3$ EPI (D) was calculated based on $\Delta B_0$ and EPI acquisition parameters in Table \ref{table1}. \label{fig5}}
\end{figure*}

\subsection{Fast brain segmentation to determine the shimming area}
$B_0$ shimming in brain tissues has been shown to effectively reduce local image distortion in brain tissues, compared to $B_0$ shimming in the whole head regions which include skulls and other non-brain tissues. \cite{Xu2023} Deep-learning-based brain extraction \cite{Shanbhag2019} can significantly reduce the computation time, enabling in-scanner implementation of automated and personalized shimming ROI. In this work, a customized deep-learning-based brain segmentation tool \cite{Shanbhag2019} was applied to segment brain tissue from the magnitude image of $B_0$ field mapping (Figure \ref{fig1}). 

The segmentation and computation of $B_0$ shimming coefficients were implemented in the scanner. The processing pipeline was fully automated and took less than 10 seconds. 

\subsection{Image acquisition}
A brain-shaped phantom filled with nickel chloride and copper sulfate and the mini-ACR (American College of Radiology) phantom were scanned. Furthermore, three healthy volunteers (2 male, 49/38 years old; 1 female, 34 years old) were recruited and scanned under a local institutional review board-approved protocol. MRI scanning was performed in a head-only investigational MAGNUS 3.0T MRI system operating at maximum performance of 300 mT/m and 750 T/m/s. \cite{foo_highly_2020} A 32-channel phased-array receive radiofrequency coils (Nova Medical, USA) was used for signal receiving. 

For each scan session, a first axial $B_0$ field map (i.e., 'Original $\Delta B_0$') covering the whole phantom or the whole brain (Figure \ref{fig1}A) was measured with 2D multi-echo spoiled gradient echo acquisition. Two sets of shimming coefficients were calculated from the acquired $B_0$ field map, i.e., (1) dynamic 2D slice-by-slice shimming assuming linear gradient fields and (2) dynamic 2D slice-by-slice shimming using actual gradient fields. The actual gradient field of each voxel was approximated by using the $10^{th}$ order spherical harmonics.\cite{Tan2013MRM} Another two $B_0$ field maps (i.e., 'Measured post-shim $\Delta B_0$, linear' and 'Measured post-shim $\Delta B_0$, actual') were measured by applying the shimming coefficients for each slice in $B_0$ field mapping using a modified 2D multi-echo spoiled gradient echo sequence with dynamic slice-by-slice shimming. Same imaging parameters were applied for all the $B_0$ field shimming, as shown in Table \ref{table1}.

Data of axial diffusion-weighted imaging (DWI) and diffusion tensor imaging (DTI) with single-shot and two-shot EPI acquisitions were acquired. The diffusion MRI sequence was modified to apply dynamic $B_0$ shimming for each slice. PGSE and OGSE diffusion encoding at 55 Hz were acquired. For each diffusion encoding, 2 x 2 x 2 $mm^3$ and 1 x 1 x 2 $mm^3$ spatial resolutions were acquired. 55 Hz was chosen as it is the maximum OGSE frequency at b-value of 1000 $s/mm^2$ and TE $\leq$ 65 ms for 1 x 1 x 2 $mm^3$ resolution. The imaging parameters are detailed in Table \ref{table1}.

In addition, $T_2$-weighted fast spin echo (FSE) anatomical image was obtained as anatomical reference. Imaging parameters include: field-of-view of 22 cm, 2 x 2 x 2 $mm^3$ and 1 x 1 x 2 $mm^3$ resolutions, and TE of 102 ms.
 
\subsection{Data Analysis} 

$B_0$ field maps after applying different shimming coefficients were also simulated by subtracting the $B_0$ shimming field and center frequency from the 'Original $\Delta B_0$'. The $B_0$ shimming field and center frequency were simulated by multiplying the actual gradient field with the shimming coefficients calculated by assuming actual gradient field (i.e., 'Predicted post-shim $\Delta B_0$, actual') or the shimming coefficients calculated by assuming linear gradient field (i.e., 'Predicted post-shim $\Delta B_0$, linear'). Finally, voxel displacement maps of EPI were calculated for comparison using the following equation:

\begin{equation} \label{eq5}
\text{Voxel displacement} =  \frac{\Delta B_0 \times \text{echo-spacing} \times \text{field-of-view}}{\text{number-of-shots}}
\end{equation}

Each of the diffusion MRI datasets were reconstructed using a customized deep learning phase correction technique,\cite{wang_ismrm_2025,wang_ismrm_2024,wang_ismrm_2023,lan_ismrm_2024} integrated with a commercialized deep learning model for denoising and deringing.\cite{lebel_performance_2020} Unlike conventional methods of using low pass filter-based approaches to estimate shot-to-shot phase variations, deep learning phase correction generates a high-quality phase with high resolution and high SNR for phase correction by using a deep learning-based network, which is a residual U-net with 4.4 million trainable parameters in approximately 10,000 kernels. The neural network was pre-trained using supervised learning from a database of over 10,000 images with various signal-to-noise levels and frequencies of background phases. Deep learning phase correction has been shown to improve robustness of complex averaging, reduce noise floor, and improve quantitative accuracy in diffusion imaging.\cite{wang_ismrm_2025,wang_ismrm_2024,wang_ismrm_2023,lan_ismrm_2024}

Diffusion-weighted images were registered to the $T_2$ weighted image (i.e., b=0 $s/mm^2$) to correct for residual diffusion gradient-induced eddy currents.\cite{Bhushan:2012, Bhushan2015} In DTI, mean diffusivity (MD), axial diffusivity (AD), radial diffusivity (RD), and fractional anisotropy (FA) were calculated after correcting for gradient nonlinearity-induced spatially varying b-value and diffusion tensor on diffusion encoding.\cite{Tan2013MRM}

\section{Results}\label{sec3}

\subsection{The impact of gradient nonlinearity on dynamic slice-by-slice $B_0$ shimming}

In the brain-shaped phantom, Figure \ref{fig2} shows $B_0$ maps of a representative slice, $\Delta B_0$ of each slice, and the histogram of corresponding voxel displacement predicted from $\Delta B_0$. Relatively large values and spatial variation of $\Delta B_0$ were presented in the 'Original $\Delta B_0$', particularly in regions away from the iso-center (Figure \ref{fig2}A). The predicted (Figures \ref{fig2} B and C) and measured (Figures \ref{fig2} D and E) post-shim $B_0$ maps both showed effective reduction of $\Delta B_0$ by using the dynamic shimming. Quantitatively, mean measured $\Delta B_0$ of each slice was reduced from -10 Hz - 25 Hz in the original $\Delta B_0$ to within $\pm$ 5 Hz by using dynamic shimming. The standard deviation and root-mean-square of $\Delta B_0$ of each slice were also reduced by a maximum of 40 Hz after applying the dynamic shimming. 

The $\Delta B_0$ maps, and mean, standard deviation, and root-mean-square of $\Delta B_0$ in slices from -80 mm (inferior) to 20 mm (superior) were consistent between 'Post-shim $\Delta B_0$, actual' and 'Post-shim $\Delta B_0$, linear', as shown in Figures \ref{fig2}G-I. Supplementary Figure S1 shows the voxel displacement in EPI acquisitions based on the $\Delta B_0$ and imaging parameters in Table \ref{table1}. In addition, the measured post-shim $\Delta B_0$ was also consistent with the predicted post-shim $\Delta B_0$. In slices further away from -80 mm, the standard deviation and root-mean-square of the 'Measured post-shim $\Delta B_0$, actual' of each slice (green hollow circles in Figures \ref{fig2} H and I) were reduced by up to 7 Hz, compared to the 'Measured post-shim $\Delta B_0$, linear' (blue hollow circles in Figures \ref{fig2} H and I). Figures \ref{fig2} J-L shows the histogram of voxel displacement of the phantom in the five most inferior slices at -110 mm to -102 mm where gradient nonlinearity is pronounced. Both 'Measured post-shim $\Delta B_0$, linear' and 'Measured post-shim $\Delta B_0$, actual' were shifted towards 0 Hz, compared to the 'Original $\Delta B_0$'. Furthermore, the histogram of 'Measured post-shim $\Delta B_0$, actual' became narrower compared to 'Measured post-shim $\Delta B_0$, linear', indicating reduced spatial variation of $\Delta B_0$. However, the measured post-shim $\Delta B_0$ (hollow circles in Figures \ref{fig2} H and I) deviated from the predicted post-shim $\Delta B_0$ (solid circles in Figures \ref{fig2} H and I).  

In human brain data of a healthy volunteer, Figure \ref{fig3} shows $B_0$ maps of a representative slice, $\Delta B_0$ of each slice, and the histogram of corresponding voxel displacement predicted from $\Delta B_0$. Similar to the phantom, dynamic $B_0$ field shimming effectively reduced the mean and root-mean-square of $\Delta B_0$ in each slice, compared to static $B_0$ field shimming (Figures \ref{fig3}G-I). The 'Measured post-shim $\Delta B_0$' (Figures \ref{fig3}D and E, and hollow circles in G-I) was consistent with the 'Predicted post-shim $\Delta B_0$' (Figures \ref{fig3}B and C, and solid circles in G-I). Unlike the phantom, the 'Post-shim $\Delta B_0$, actual' did not show significant difference to 'Post-shim $\Delta B_0$, linear', as indicated in the $B_0$ maps (Figure \ref{fig3}B-E), the mean, standard deviation, and root-mean-square of $\Delta B_0$ of each slice (Figure \ref{fig3}G-I), and the histogram of voxel displacement in the brain at inferior locations (Figures \ref{fig3} K and L). 

\subsection{Dynamic slice-by-slice $B_0$ shimming in single-shot and multi-shot EPI}

In the DWI and DTI data acquisitions with dynamic slice-by-slice $B_0$ field shimming, actual gradient field was used in the estimation of shimming coefficients. 

Figure \ref{fig4} shows the $B_0$ maps, $T_2$-weighted EPI at 2 x 2 x 2 $mm^3$ and 1 x 1 x 2 $mm^3$ resolutions, and $T_2$ FSE of representative slices of the mini-ACR phantom in axial (Figure \ref{fig4}A) and sagittal (Figure \ref{fig4}B) views, as well as the mean, standard deviation, and root-mean-square of voxel displacement of each axial slice in single-shot and two-shot EPI acquisitions (Figures \ref{fig4}C and D). By acquiring single-shot EPI with static shimming, image shift and distortion were observed, as indicated in regions highlighted by red circles and rectangles (Figure \ref{fig4}A). For example, the maximum voxel displacement of the hollow rectangle compared to $T_2$ FSE, as indicated by the blue lines in Figure \ref{fig4}A, was 4.26 mm in 2 x 2 x 2 $mm^3$ and 7.01 mm in 1 x 1 x 2 $mm^3$ images. Both dynamic shimming and two-shot EPI effectively reduce image distortion. The voxel displacement of the hollow rectangle was reduced to 1.62 mm in 2 x 2 x 2 $mm^3$ and 2.92 in 1 x 1 x 2 $mm^3$ in two-shot EPI with dynamic shimming. Furthermore, by applying dynamic slice-by-slice shimming, both image shift and distortions were further reduced for each slice, as demonstrated in the sagittal reformat (Figure \ref{fig4}B). The mean and root-mean-square of voxel displacement of each slice in single-shot EPI acquisition with dynamic shimming were reduced by a maximum of 10 voxels in 2 x 2 x 2 $mm^3$ image (Figure \ref{fig4}C) and 20 voxels in 1 x 1 x 2 $mm^3$ image (Figure \ref{fig4}D), compared to single-shot EPI acquisition with static shimming. Two-shot EPI acquisitions with dynamic shimming also achieved reduced mean and root-mean-square of voxel displacement of each slice by a maximum of 5 and 10 voxels in the two resolution images, compared to two-shot EPI acquisitions with static shimming. 

Figure \ref{fig5} shows the $B_0$ maps, $T_2$-weighted EPI at 2 x 2 x 2 $mm^3$ and 1 x 1 x 2 $mm^3$ resolutions, and $T_2$ FSE of representative slices of a healthy volunteer's brain in axial (Figure \ref{fig5}A) and coronal (Figure \ref{fig5}B) views, as well as the mean, standard deviation, and root-mean-square of voxel displacement of each axial slice in single-shot and two-shot EPI acquisitions (Figures \ref{fig5}C and D). Image distortions were more apparent in images with 1 x 1 x 2 $mm^3$ resolution, due to the longer echo spacing. Applying dynamic slice-by-slice shimming more apparently reduced distortions of images with 1 x 1 x 2 $mm^3$ resolution, especially in slices at the bottom of the brain, as indicated by the green contours and red arrows. Supplementary Figures S2 and S3 showed the voxel displacement of each axial slice in the other two healthy volunteers. In all three volunteers, dynamic slice-by-slice $B_0$ field shimming more efficiently reduced root-mean-square of voxel displacement at the edge of the inferior brain regions. In one subject (Supplementary Figure S2), dynamic slice-by-slice $B_0$ shimming also reduced voxel displacement by a maximum of 3 voxels at 2 x 2 x 2 $mm^3$ resolution and a maximum of 6 voxels at 1 x 1 x 2 $mm^3$ resolution from slices at -40 mm to 40 ms.

\subsection{Human brain OGSE and PGSE diffusion MRI with brain tissue-selective dynamic slice-by-slice $B_0$ shimming}

\begin{figure*}
\centerline{\includegraphics[width=400pt,height=52pc]{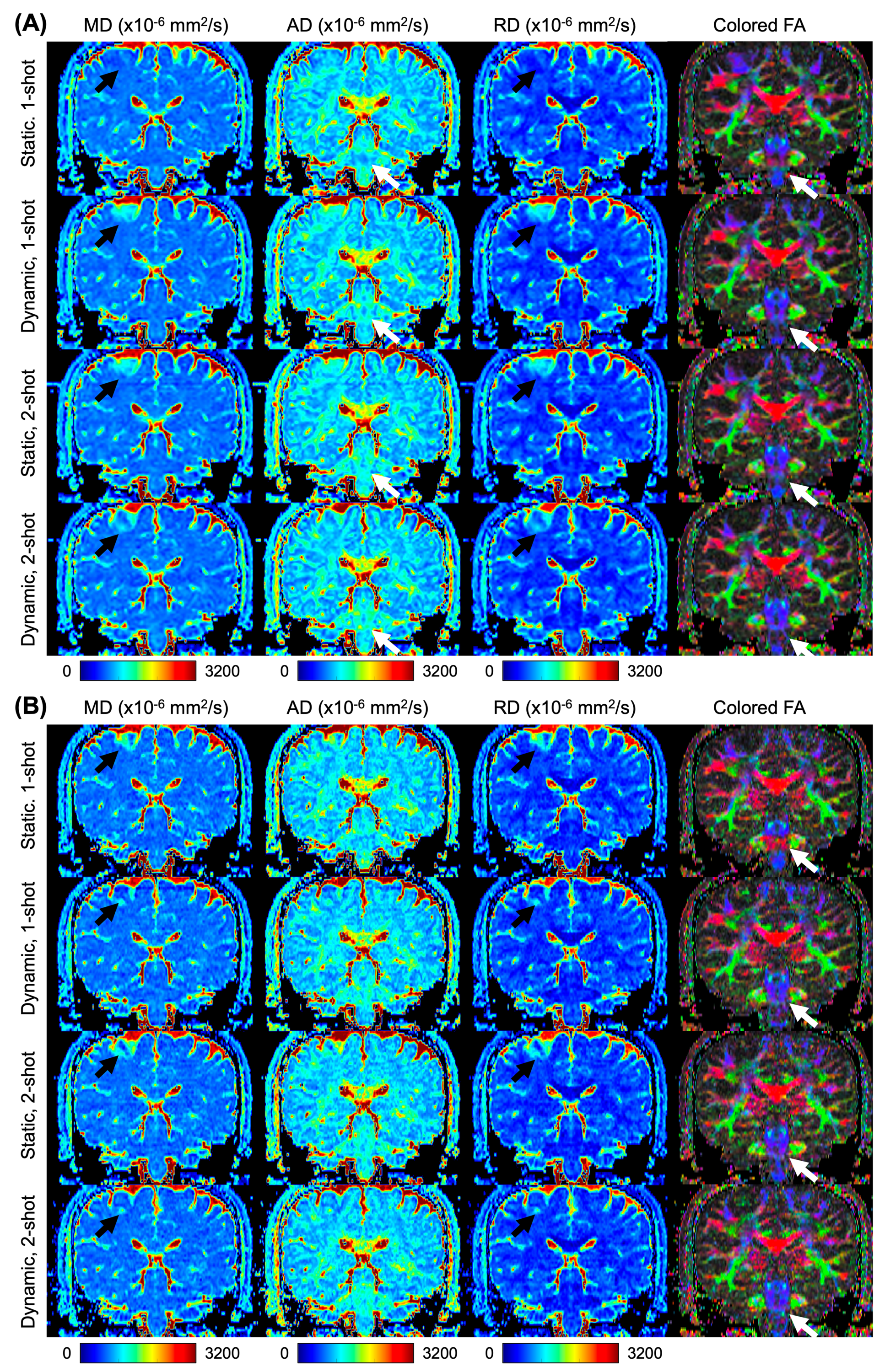}}
\caption{MD/AD/RD/Colored FA in coronal reformat of PGSE DTI acquired at 2 x 2 x 2 $mm^3$ resolution (A) and 1 x 1 x 2 $mm^3$ (B) resolution using different shimming approaches. The images were from the same subject as shown in Figure \ref{fig5} (i.e., Subject \#2). Black arrows and white arrows point to the inaccurate estimation of MD/AD/RD/FA in the superior region of the brain and the brainstem, respectively, acquired with single-shot EPI with static $B_0$ field shimming. Dynamic $B_0$ field shimming and two-shot EPI acquisitions both improve the accuracy of diffusion measurements. \label{fig6}}
\end{figure*}

Figure \ref{fig6} shows the MD/AD/RD/FA of PGSE DTI of a healthy volunteer acquired at 2 x 2 x 2 $mm^3$ (Figure \ref{fig6}A) and 1 x 1 x 2 $mm^3$ (Figure \ref{fig6}B) resolutions using different $B_0$ field shimming approaches. The brainstem showed similar AD and FA across dynamic 1-shot acquisition, static 2-shot acquisition, and dynamic 2-shot acquisition, which differ from AD and FA estimated from the static 1-shot acquisition, as pointed by the white arrows. Cerebral cortex and white matter at the superior regions of the brain also showed similar MD, AD,RD, and FA across dynamic 1-shot acquisition, static 2-shot acquisition, and dynamic 2-shot acquisition, which differ from that estimated from the static 1-shot acquisition, as pointed by the black arrows. The effect of $B_0$-induced image shift and distortion on the diffusivity is more discernible at 1 x 1 x 2 $mm^3$ resolution, compared to 2 x 2 x 2 $mm^3$. For example, the colored FA in the posterior brain regions (as pointed by the white arrows in Figure \ref{fig6}B) showed blue color in dynamic 1-shot acquisition, static 2-shot acquisition, and dynamic 2-shot acquisition, indicating white matter tracts in the brainstem along superior-inferior directions. In comparison, the colored FA in the same regions in static 1-shot acquisition showed red color, indicating white matter tracts in the cerebral peduncle which was misplaced in the image due to large $\Delta B_0$. 

\begin{figure*}
\centerline{\includegraphics[width=500pt,height=16pc]{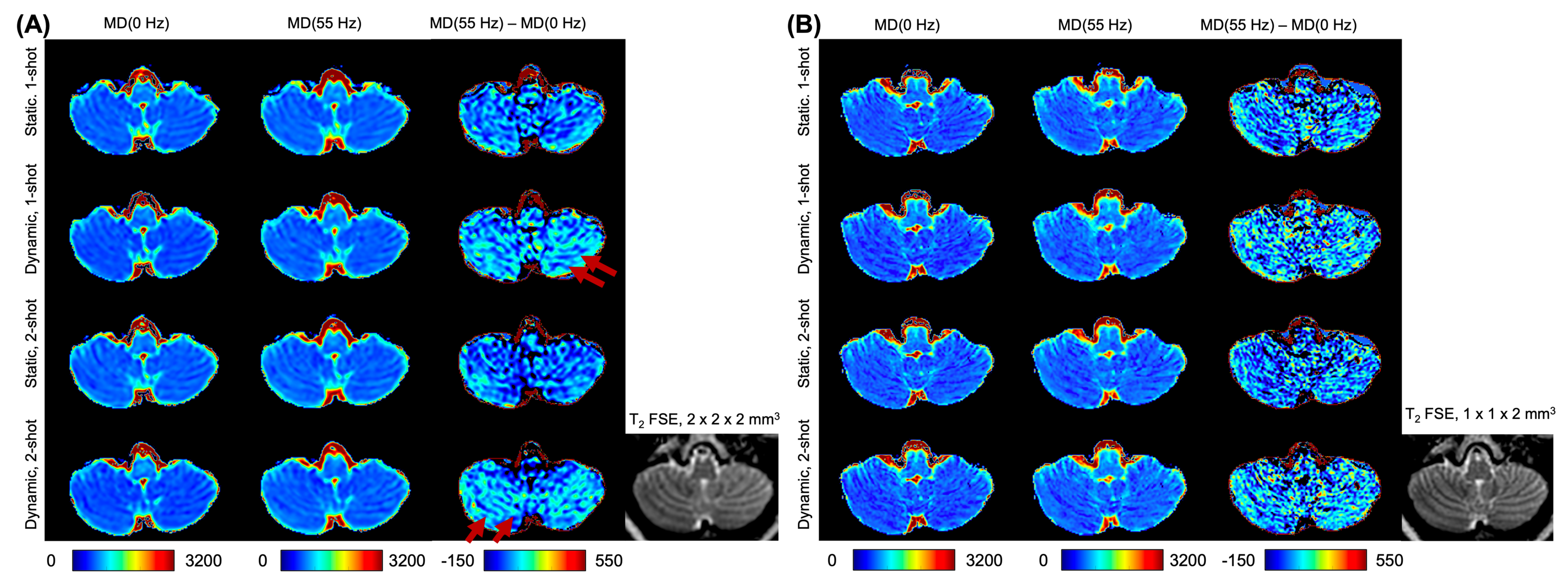}}
\caption{MD(0 Hz) acquired using PGSE, MD(55 Hz) acquired using OGSE, and the difference between the two MD's in a representative axial slice covering the cerebellum of the healthy volunteer shown in Figures \ref{fig5} and \ref{fig6} (i.e., Subject \#2). Images in (A) were acquired with 2 x 2 x 2 $mm^3$ resolution and images in (B) were acquired with 1 x 1 x 2 $mm^3$ resolution. \label{fig7}}
\end{figure*}

Figure \ref{fig7} shows the axial view of cerebellum MD from the PGSE DTI (approximated as MD(0 Hz)) and the OGSE DTI at 55 Hz (i.e., MD(55 Hz)), and the difference between MD(55 Hz) and MD(0 Hz) at 2 x 2 x 2 $mm^3$ (Figure \ref{fig7}A) and 1 x 1 x 2 $mm^3$ (Figure \ref{fig7}B) resolutions. Cerebellar white matter (i.e., hypo-intensity regions in the $T_2$ FSE image) showed relatively high MD(0 Hz) and MD(55 Hz) acquired from all shimming approaches, whereas cerebellar grey matter (i.e., hyper-intensity regions in the $T_2$ FSE image) showed relatively low MD(0 Hz) and MD(55 Hz). In the MD difference map, more hyper-intensity regions that correspond to the cerebellar white matter regions in the $T_2$ FSE image (red arrows in Figure \ref{fig7}) were observed using the dynamic $B_0$ field shimming, compared to that using the static $B_0$ field shimming.

\section{Discussions and Conclusions}\label{sec4}

We have proposed a brain tissue selective, gradient nonlinearity-informed optimization of dynamic slice-by-slice $B_0$ field shimming to reduce $B_0$ field inhomogeneity in human head-only high-performance gradient 3T MRI systems. We have demonstrated that incorporating the actual gradient field of the X, Y, and Z gradient coils in regions with high gradient nonlinearity can improve $B_0$ field shimming in a brain-shaped phantom. Furthermore, we have demonstrated that dynamic slice-by-slice $B_0$ field shimming results in reduced image distortion and benefits brain diffusion MRI, especially for high-resolution imaging (e.g., by a maximum root-mean-square of 6 voxel shifts at 1 x 1 x 2 $mm^3$ resolution). The proposed method will greatly improve the image quality of high-resolution advanced brain diffusion MRI that has emerged for tissue microstructure imaging at 3T,\cite{Setsompop2013, Weiger2018MRM, foo_highly_2020} 7T, \cite{Feinberg2023} and 10.5T.\cite{rutt2025}

We have preliminarily demonstrated that $T_2$ FSE hypo-intensity regions in the cerebellum showed higher difference between MD at OGSE 55 Hz and MD at PGSE by acquiring diffusion MRI with dynamic $B_0$ field shimming, compared to that with static $B_0$ field shimming. Wu et al \cite{Wu2014} has also shown $T_2$ FSE hypo-intensity regions, including the cerebellar granule cell layer, presenting high difference between ADC at OGSE 220 Hz and ADC at PGSE in the cerebellum of adult rats. Our study highlighted the potential to translate the characterization of fine layers in the cerebellum using high-resolution OGSE diffusion MRI with dynamic $B_0$ field shimming in humans. We acknowledge that the signal-to-noise ratio in MD(55 Hz)-MD(0 Hz) may not be sufficient for quantitative evaluation in this study. This is because we covered the whole brain for diffusion MRI to show the effectiveness of dynamic slice-by-slice $B_0$ shimming, which unfortunately limits the number of diffusion encoding directions in a feasible scan time. Future studies will optimize the imaging protocol, including the number of diffusion encoding directions, b-value, OGSE frequencies, and TE, to comprehensively characterize the time-dependent diffusion of the human cerebellum.

In the phantom, we have demonstrated that incorporating the gradient nonlinearity in the calculation of shimming coefficients can further reduce $\Delta B_0$ where gradient nonlinearity is not negligible. However, the measured post-shim $\Delta B_0$ deviate from the predicted post-shim $\Delta B_0$, especially at slices further away from the iso-center. In this study, the actual gradient field was approximated using the $10^{th}$ order spherical harmonics of the electromagnetic field of the designed gradient coils, which may differ from the actual gradient field of the manufactured  coils. Using a field camera may potentially map the actual gradient field of the manufactured coils, which will be studied in the future. Other potential reasons include the error on the applied current for the first-order shimming due to digitization.

Incorporating gradient nonlinearity in $B_0$ field shimming did not show significant reduction of mean and standard deviation of $\Delta B_0$ in the cerebellum in the human brain. In human brains, the tissue susceptibility-induced $B_0$ field inhomogeneity is composed of spatially high-order terms. Although the gradient nonlinearity is unwanted for k-space encoding and gradient-induced image contrast (e.g., diffusion encoding), the nonlinear gradient might more effectively reduce the $B_0$ field inhomogeneity compared to a perfectly linear gradient. This depends upon whether the actual gradient field of gradient coils has similar spatial distribution as tissue susceptibility-induced $B_0$ field. Future works will compare the effectiveness of $B_0$ field shimming in the brain with high linearity, e.g., in whole-body MRI systems, versus that with low linear gradients, e.g., in head-only MRI systems. 

Furthermore, phantom results showed that single-shot EPI acquisition with dynamic slice-by-slice $B_0$ field shimming may achieve similar reduced image distortion without increasing scan time, compared to that of two-shot EPI acquisition. Notably, in slices farther from the iso-center, single-shot EPI acquisition with dynamic shimming even outperforms two-shot with static shimming, based on the root-mean-square of $\Delta B_0$. In human brains, however, single-shot EPI acquisition with dynamic shimming did not achieve similarly small root-mean-square of $\Delta B_0$ in every slice of all the three subjects as that in the two-shot EPI acquisition with static shimming. This is likely due to the greater presence of high-order $B_0$ terms in vivo due to . Therefore, multi-shot EPI acquisitions are still superior and robust in reducing image distortions caused by $\Delta B_0$, regardless the spatial orders of the $\Delta B_0$ field. Nonetheless, we observed that combining two-shot EPI with dynamic shimming achieves the least distortion and yields the most accurate spatial representation compared to $T_2$ FSE anatomical images. Therefore, the use of dynamic shimming in two-shot EPI, especially at high spatial resolution such as 1 mm in-plane, is still important.

The proposed $B_0$ field shimming calculation incorporating the actual gradient field can benefit other applications and other MRI systems including whole-body systems and specialized local gradient coils.\cite{Bhuiyan2021} For example, $B_0$ field shimming is also key to MR spectroscopy. In single-voxel MR spectroscopy, $B_0$ field shimming is optimized for the prescribed voxel, which can also be at off-center locations. Furthermore, in body applications in the whole-body MRI system, the effect of gradient nonlinearity of the X and Y gradient axes may be substantial when the volume-of-interest is prescribed in the liver, breast, and other organs at the off-center locations along the left-right and anterior-posterior directions. Therefore, $B_0$ field shimming can be further improved by incorporating the actual gradient field. 

Additionally, the dynamic slice-by-slice $B_0$ shimming technique can also benefit other gradient-echo and spin-echo applications with EPI and spiral readout trajectories, e.g. arterial spin labeling,\cite{Ji2025} functional MRI, and MR elastography at 3T, 5T, and ultrahigh field. In gradient-echo functional MRI, in addition to the reduction of distortion or blurring, dynamic shimming could also mitigate signal dropout due to $B_0$ field inhomogeneities, especially in regions with high susceptibility differences, e.g. the orbitofrontal cortex. In arterial spin labeling, dynamic shimming between the labeling and imaging can improve labeling efficiency. 

This study has several limitations. First, the voxel displacement was calculated from the measured $\Delta B_0$ from $B_0$ mapping, which may not reflect the overall voxel displacement in EPI images which is not only caused by $B_0$ field inhomogeneity, but also eddy currents of EPI gradients. Second, the advantage of slice-by-slice $B_0$ shimming in simultaneous multislice EPI acquisitions may be limited due to the difficulty in reducing $\Delta B_0$ in non-adjacent slices.\cite{Setsompop2012} The potential advantage and further optimization of dynamic $B_0$ field shimming in simultaneous multislice EPI acquisitions \cite{Hetherington2021} will be studied in the future. In addition, the slice-by-slice $B_0$ shimming may not effectively reduce $\Delta B_0$ in 3D imaging, which may be best addressed by high-order static $B_0$ shimming. Furthermore, we have shown that dynamic slice-by-slice $B_0$ field shimming with zeroth and first order outperforms high-order static $B_0$ field shimming with up to the second order in breast MRI.\cite{Zhu2025ismrm} It would be interesting to compare the effectiveness of $\Delta B_0$ reduction between the two shimming approaches in human brains. This will be studied in human head-only high-performance gradient MRI systems with high-order shimming coils and drivers in the future.

In conclusion, diffusion MRI with brain tissue-selective, dynamic slice-by-slice $B_0$ effectively reduces image distortion and improves the accuracy of diffusivity characterization in high-resolution images. The impact of the proposed technique on diffusion MRI-based tractography and time-dependent diffusion MRI will be studied in in vivo human to understand the function and connectivity of human brain, especially in different layers in the human cerebellum.

%\filbreak

\section*{Acknowledgments}
Dr. Lan contributed to the pulse sequence, tools, data acquisition and analysis, and the manuscript drafting and revision. Dr. Sherry Huang contributed to the software, data analysis, and the manuscript revision. Dr. Bhushan contributed to the software and the manuscript revision. Dr. Wang contributed to the data analysis and the manuscript revision. Dr. Lee contributed to the theory and the manuscript revision. Dr. Raymond Huang, Dr. Maller, and Dr. McNab contributed to the data analysis and the manuscript revision. Dr. Zhu contributed to the theory, experiment design, tools, data acquisition and analysis, and the manuscript drafting and revision.

The authors thank Dr. Gaohong Wu for helping on pulse sequence modification and Mr. Eric Fiveland for support with MRI scans. 

Research reported in this publication was supported by the National Institute of Biomedical Imaging and Bioengineering of the National Institutes of Health (NIH) under award Number U01EB034313. The content is solely the responsibility of the authors and does not necessarily represent the official views of the NIH. 

\subsection*{CONFLICT OF INTEREST STATEMENT}

Dr. Lan, Dr. Sherry Huang, Dr. Bhushan, Dr. Wang, Dr. Lee, and Dr. Zhu are employees of GE HealthCare.

\bibliography{MRM-AMA}%

\vspace*{6pt}
%\appendix

\vfill\pagebreak

\end{document}